\documentclass[11pt,a4paper]{article}
\usepackage[hyperref]{acl2020}
\usepackage{times}
\usepackage{latexsym}

\usepackage{graphicx}
\usepackage{amsmath,amssymb}
\usepackage{booktabs}
\usepackage{xspace}
\usepackage{amsthm}
\usepackage{listings}

\usepackage{amssymb}
\usepackage{pifont}
\usepackage{bbm}
\newcommand{\cmark}{\ding{51}}%
\newcommand{\xmark}{\ding{55}}%
\usepackage[inline]{enumitem}
\usepackage{float}

\usepackage{microtype}

\newcommand\la{$\langle$}
\newcommand\ra{$\rangle$}

\usepackage{url}

\aclfinalcopy %
\def\aclpaperid{1923} %

\newcommand\jh[1]{}
\newcommand\tm[1]{}

\newcommand\sqa{\textsc{SQA}\xspace}
\newcommand\wikisql{\textsc{WikiSQL}\xspace}
\newcommand\wtq{\textsc{WikiTQ}\xspace}

\newcommand{\ours}{\textsc{TaPas}\xspace}

\definecolor{myred}{RGB}{168, 5, 5}
\definecolor{mygreen}{RGB}{0, 128, 11}

\definecolor{dgreen}{RGB}{20,80,20}

\title{\ours : Weakly Supervised Table Parsing via Pre-training}

\author{Jonathan Herzig\textsuperscript{1,2}, Pawe\l{} Krzysztof Nowak\textsuperscript{1}, Thomas M{\"u}ller\textsuperscript{1},\\ \textbf{Francesco Piccinno\textsuperscript{1}, Julian Martin Eisenschlos\textsuperscript{1}} \\
\\
  \textsuperscript{1}Google Research \\
  \textsuperscript{2}School of Computer Science, Tel-Aviv University \\
  {\tt \{jherzig,pawelnow,thomasmueller,piccinno,eisenjulian\}@google.com} \\}

\date{}

\begin{document}
\maketitle

\begin{abstract}
  Answering natural language questions over tables is usually seen as a semantic parsing task. 
To alleviate the collection cost of full logical forms,
one popular approach
focuses on
weak supervision consisting of denotations instead of logical forms. However, training semantic parsers from weak supervision poses difficulties, and in addition, the generated logical forms are only used as an intermediate step prior to retrieving the denotation.
In this paper, we present \ours, an approach to question answering over tables without generating logical forms. \ours trains from weak supervision, and predicts the denotation 
by selecting table cells and optionally applying a corresponding aggregation operator to such selection.
\ours extends BERT's architecture to encode tables as input, initializes from an effective joint pre-training of text segments and tables crawled from Wikipedia, and is trained end-to-end.   
We experiment with three different semantic parsing datasets, and find that \ours outperforms or rivals semantic parsing models by improving state-of-the-art accuracy on \sqa{} from $55.1$ to $67.2$ and performing on par with the state-of-the-art on \wikisql{} and \wtq{}, but with a simpler model architecture. We additionally find that transfer learning, which is trivial in our setting, from \wikisql{} to \wtq{}, yields $48.7$ accuracy, $4.2$ points above the state-of-the-art. 
\end{abstract}

\section{Introduction}

Question answering from semi-structured tables is usually seen as a semantic parsing task where the question is translated to a logical form that can
be executed against the table to retrieve the correct denotation \cite{pasupat2015compositional, zhong2017seq2sql, dasigi2019iterative, agarwal2019learning}. Semantic parsers rely on supervised training data that pairs natural language questions with logical forms, but such data is expensive to annotate. 

In recent years, many attempts aim to reduce the burden of data collection for semantic parsing, including paraphrasing \cite{wang2015overnight}, human in the loop \cite{iyer2017neural,lawrence-riezler-2018-improving} and training on examples from other domains \cite{herzig2017multi,su2017cross}.
One prominent data collection approach focuses on weak supervision where a training example consists of a question and its denotation instead of the full logical form \cite{clarke10world, liang11dcs, artzi2013weakly}. Although appealing, training semantic parsers from this input is often difficult due to the abundance of spurious logical forms \cite{berant2013freebase, guu2017bridging} and reward sparsity \cite{agarwal2019learning,muhlgay2019value}.

In addition, semantic parsing applications only utilize the generated logical form as an intermediate step in retrieving the answer. Generating logical forms, however, introduces difficulties such as maintaining a logical formalism with sufficient expressivity, obeying decoding constraints (e.g. well-formedness), and the label bias problem \cite{andor2016globally,lafferty01crf}.

In this paper we present \ours (for \textbf{Ta}ble \textbf{Pa}r\textbf{s}er), a weakly supervised question answering model that reasons over tables without generating logical forms. \ours predicts a minimal program by selecting a subset of the table cells and a possible aggregation operation to be executed on top of them. Consequently, \ours can learn operations from natural language, without the need to specify them in some formalism.
This is implemented by extending BERT's architecture \cite{devlin2018BERT} with additional embeddings that capture tabular structure, and with two classification layers for selecting cells and predicting a corresponding aggregation operator.

Importantly, we introduce a pre-training method for \ours, crucial for its success on the end task. We extend BERT's masked language model objective to structured data, and pre-train the model over millions of tables and related text segments crawled from Wikipedia. During pre-training, the model masks some tokens from the text segment and from the table itself, where the objective is to predict the original masked token based on the textual and tabular context.

Finally, we present an end-to-end differentiable training recipe that allows \ours to train from weak supervision. For examples that only involve selecting a subset of the table cells, we directly train the model to select the gold subset. For examples that involve aggregation, the relevant cells and the aggregation operation are not known from the denotation. In this case, we calculate an expected soft scalar outcome over all aggregation operators given the current model, and train the model with a regression loss against the gold denotation.   

In comparison to prior attempts to reason over tables without generating logical forms \cite{neelakantan2016neural, yin-etal-2016-neural, muller2019answering}, \ours achieves better accuracy, and holds several advantages: its architecture is simpler as it includes a single encoder with no auto-regressive decoding, it enjoys pre-training, tackles more question types such as those that involve aggregation, and directly handles a conversational setting.

We find that on three different semantic parsing datasets, \ours performs better or on par in comparison to other semantic parsing and question answering models. On the conversational \sqa \cite{iyyer2017search}, \ours improves state-of-the-art accuracy from $55.1$ to $67.2$, and achieves on par performance on \wikisql \cite{zhong2017seq2sql} and \wtq \cite{pasupat2015compositional}. 
Transfer learning, which is simple in \ours, from \wikisql to \wtq achieves 48.7 accuracy, $4.2$ points higher than state-of-the-art.
Our code and pre-trained model are publicly available at \url{https://github.com/google-research/tapas}.
\section{\ours Model}
\label{sec:model}

\begin{figure}[t]
\centering
\includegraphics[width=1.0\columnwidth]{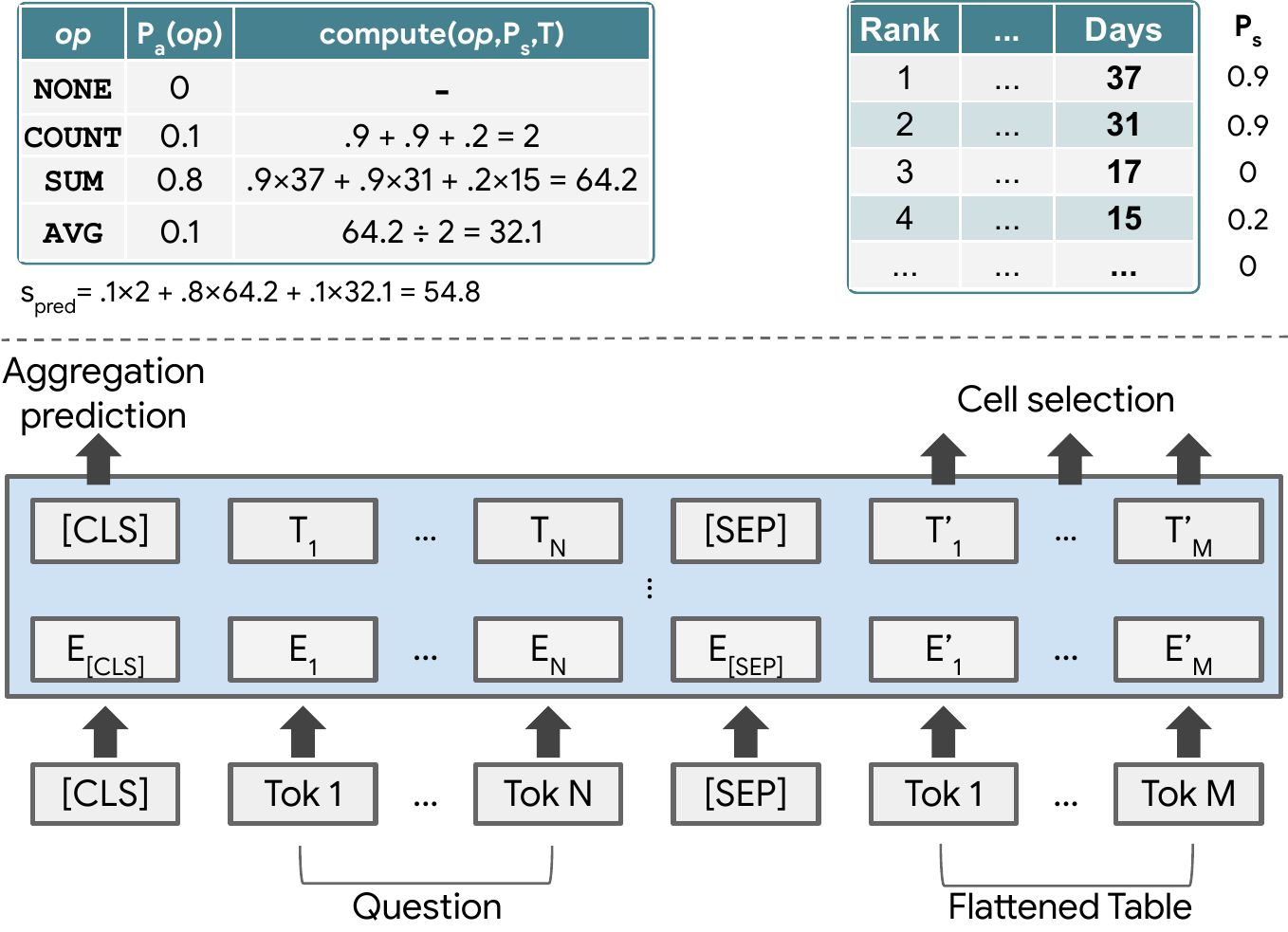}
\caption{\ours model (bottom) with example model outputs for the question: \textit{``Total number of days for the top two''}. Cell prediction (top right) is given for the selected column's table cells in bold (zero for others) along with aggregation prediction (top left).}
\label{fig:full_model}
\end{figure}

\begin{figure*}[t]
\centering
\includegraphics[width=1.0\textwidth]{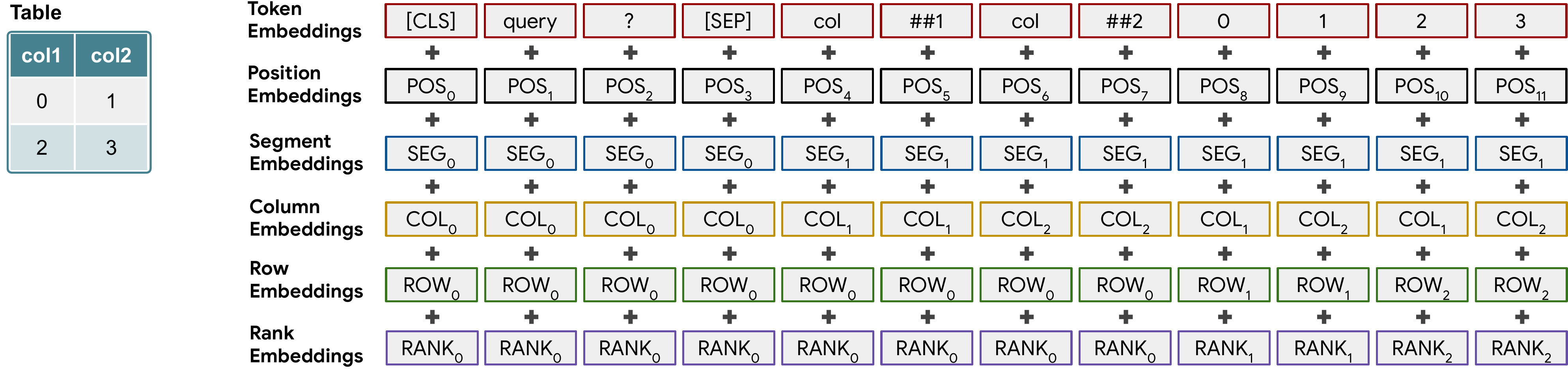}
\caption{Encoding of the question ``query?'' and a simple table using the special embeddings of \ours.
The previous answer embeddings are omitted for brevity.
}
\label{fig:input_example}
\end{figure*}

Our model's architecture (Figure \ref{fig:full_model}) is based on BERT's encoder with additional positional embeddings used to encode tabular structure (visualized in Figure \ref{fig:input_example}). We flatten the table into a sequence of words, split words into word pieces (tokens) and concatenate the question tokens before the table tokens. We additionally add two classification layers for selecting table cells and aggregation operators that operate on the cells. We now describe these modifications and how inference is performed.

\paragraph{Additional embeddings} 

We add a separator token between the question and the table, but unlike \newcite{hwang2019comprehensive} not between cells or rows. Instead, the token embeddings are combined with table-aware positional embeddings before feeding them to the model. We use different kinds of positional embeddings:

\begin{itemize}[leftmargin=*,itemsep=0pt]
    \item \textbf{Position ID} is the index of the token in the flattened sequence (same as in BERT).
    \item \textbf{Segment ID} takes two possible values: 0 for the question, and 1 for the table header and cells.
    \item \textbf{Column / Row ID} is the index of the column/row that this token appears in, or 0 if the token is a part of the question.
    \item \textbf{Rank ID} if column values can be parsed as floats or dates, we sort them accordingly and assign an embedding based on their numeric rank (0 for not comparable, 1 for the smallest item, $i+1$ for an item with rank $i$). This can assist the model when processing questions that involve superlatives, as word pieces may not represent numbers informatively \cite{wallace2019nlp}.
    \item \textbf{Previous Answer} given a conversational setup where the current question might refer to the previous question or its answers (e.g., question 5 in Figure \ref{fig:dataset_examples}), we add a special embedding that marks whether a cell token was the answer to the previous question (1 if the token's cell was an answer, or 0 otherwise).
\end{itemize}

\paragraph{Cell selection}
This classification layer selects a subset of the table cells. Depending on the selected aggregation operator, these cells can be the final answer or the input used to compute the final answer.
Cells are modelled as independent Bernoulli variables. First, we compute the logit for a token using a linear layer on top of its last hidden vector. Cell logits are then computed as the average over logits of tokens in that cell. The output of the layer is the probability $p^{(c)}_{\text{s}}$ to select cell $c$.
We additionally found it useful to add an inductive bias to select cells within a single column. We achieve this by introducing a categorical variable to select the correct column. The model computes the logit for a given column by applying a new linear layer to the average embedding for cells appearing in that column. We add an additional column logit that corresponds to selecting no column or cells. We treat this as an extra column with no cells. The output of the layer is the probability $p_{\mathrm{col}}^{(co)}$ to select column $co$ computed using softmax over the column logits. We set cell probabilities $p_s^{(c)}$ outside the selected column to $0$.

\paragraph{Aggregation operator prediction}
Semantic parsing tasks require discrete reasoning over the table, such as summing numbers or counting cells. To handle these cases without producing logical forms, \ours outputs a subset of the table cells together with an optional \textit{aggregation operator}. The aggregation operator describes an operation to be applied to the selected cells, such as \texttt{SUM}, \texttt{COUNT}, \texttt{AVERAGE} or \texttt{NONE}. The operator is selected by a linear layer followed by a softmax on top of the final hidden vector of the first token (the special \texttt{[CLS]} token). We denote this layer as $p_a(op)$, where $op$ is some aggregation operator.

\paragraph{Inference} We predict the most likely aggregation operator together with a subset of the cells (using the cell selection layer).
To predict a discrete cell selection we select all table cells for which their probability is larger than $0.5$.
These predictions are then executed against the table to retrieve the answer, by applying the predicted aggregation over the selected cells.

\section{Pre-training}
\label{sec:pre_training}

Following the recent success of pre-training models on textual data for natural language understanding tasks, we wish to extend this procedure to structured data, as an initialization for our table parsing task. To this end, we pre-train \ours on a large number of tables from Wikipedia. 
This allows the model to learn many interesting correlations between text and the table, and between the cells of a columns and their header.

We create pre-training inputs by extracting text-table pairs from Wikipedia.
We extract 6.2M tables: 3.3M of class \emph{Infobox}\footnote{\url{en.wikipedia.org/wiki/Help:Infobox}} and 2.9M of class \emph{WikiTable}.
We consider tables with at most 500 cells.
All of the end task datasets we experiment with only contain horizontal tables with a header row with column names. 
Therefore, we only extract Wiki tables of this form using the \texttt{<th>} tag to identify headers. We furthermore, transpose Infoboxes into a table with a single header and a single data row.
The tables, created from Infoboxes, are arguably not very typical, but we found them to improve performance on the end tasks.

As a proxy for questions that appear in the end tasks, we extract 
the table caption, article title, article description, segment title and text of the segment the table occurs in as relevant text snippets. In this way we extract 21.3M snippets.

We convert the extracted text-table pairs to pre-training examples as follows:
Following \newcite{devlin2018BERT}, we use a masked language model pre-training objective. 
We also experimented with adding a second objective of predicting whether the table belongs to the
text or is a random table but did not find this to improve the performance on the end tasks.
This is aligned with \newcite{roberta2019} that similarly did not benefit from a next sentence prediction task.

For pre-training to be efficient, we restrict our word piece sequence length to a certain budget (e.g., we use 128 in our final experiments). That is, the combined length of tokenized text and table cells has to fit into this budget. To achieve this, we randomly select a snippet of $8$ to $16$ word pieces from the associated text. To fit the table, we start by only adding the first word of each column name and cell. We then keep adding words turn-wise until we reach the word piece budget.
For every table we generate 10 different snippets in this way.

We follow the masking procedure introduced by BERT. We use whole word masking\footnote{\url{https://github.com/google-research/bert/blob/master/README.md}} for the text,
and we find it beneficial to apply \textit{whole cell masking} (masking all the word pieces of the cell if any of its pieces is masked) to the table as well.

We note that we additionally experimented with data augmentation, which shares a similar goal to pre-training. We generated synthetic pairs of questions and denotations over real tables via a grammar, and augmented these to the end tasks training data. As this did not improve end task performance significantly, we omit these results.

\section{Fine-tuning}
\label{sec:fine_tuning}

\begin{figure*}[t]
\centering
\includegraphics[width=1.0\textwidth]{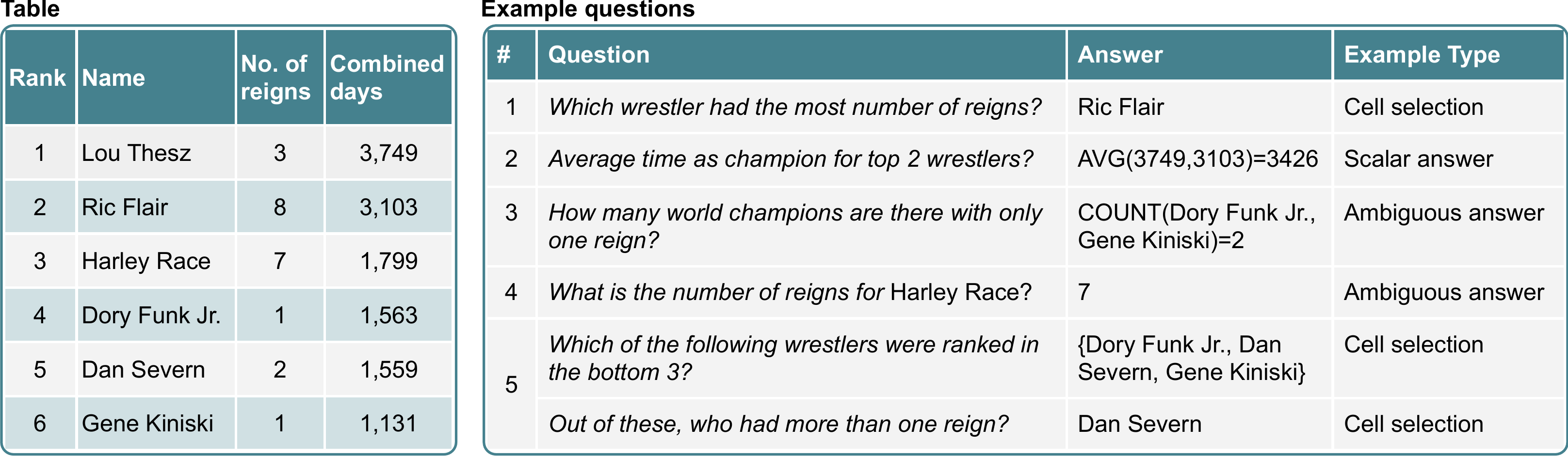}
\caption{A table (left) with corresponding example questions (right). The last example is conversational.}
\label{fig:dataset_examples}
\end{figure*}

\paragraph{Overview}
 
We formally define table parsing in a weakly supervised setup as follows. Given a training set of $N$
examples $\{(x_i,T_i,y_i)\}_{i=1}^{N}$, where $x_i$ is an utterance, $T_i$ is a table and $y_i$ is a corresponding set of denotations, our goal is to learn a model that maps a new utterance $x$ to a program $z$, such that when $z$ is executed against the corresponding table $T$, it yields the correct denotation $y$. The program $z$ comprises a subset of the table cells and an optional aggregation operator. The table $T$ maps a table cell to its value.

As a pre-processing step described in Section~\ref{sec:datasets}, we translate the set of denotations $y$ for each example to a tuple $(C, s)$ of cell coordinates $C$ and a scalar $s$, which is only populated when $y$ is a single scalar.
We then guide training according to the content of $(C, s)$. For \textit{cell selection} examples, for which $s$ is not populated, we train the model to select the cells in $C$. For \textit{scalar answer} examples, where $s$ is populated but $C$ is empty, we train the model to predict an aggregation over the table cells that amounts to $s$. We now describe each of these cases in detail. 

\paragraph{Cell selection} In this case $y$ is mapped to a subset of the table cell coordinates $C$ (e.g., question 1 in Figure \ref{fig:dataset_examples}). For this type of examples, we use a hierarchical model that first selects a single column and then cells from within that column only. 

We directly train the model to select the column $\textrm{col}$ which has the highest number of cells in $C$. For our datasets cells $C$ are contained in a single column and so this restriction on the model provides a useful inductive bias. If $C$ is empty we select the additional empty column corresponding to empty cell selection.
The model is then trained to select cells $C \cap \textrm{col}$ and not select $(T \setminus C) \cap \textrm{col}$. The loss is composed of three components: (1) the average binary cross-entropy loss over column selections:
\begin{align*}
\mathcal{J}_{\text{columns}} &= \frac{1}{\mathopen|\mathrm{Columns}\mathclose|} \sum_{\mathrm{co} \in \mathrm{Columns}} \mathrm{CE}(p_\textrm{col}^{(\textrm{co})}, \mathbbm{1}_{\textrm{co} = \textrm{col}})
\end{align*}
where the set of columns $\textrm{Columns}$ includes the additional empty column, $\mathrm{CE}(\cdot)$ is the cross entropy loss, $\mathbbm{1}$ is the indicator function.
(2) the average binary cross-entropy loss over column cell selections:
\begin{align*}
  \mathcal{J}_{\text{cells}} &= \frac{1}{\mathopen|\textrm{Cells}(\textrm{col})\mathclose|} \sum_{c \in \textrm{Cells}(\textrm{col})} \mathrm{CE}(p^{(c)}_{\text{s}},\mathbbm{1}_{c \in C}),
\end{align*}
where $\textrm{Cells}(\textrm{col})$ is the set of cells in the chosen column.
(3) As for \textit{cell selection} examples no aggregation occurs, we define the aggregation supervision to be \texttt{NONE} (assigned to $op_0$), and the aggregation loss is:
\begin{equation*}
  \mathcal{J}_{\text{aggr}} = -\log p_{\text{a}}(op_0) .
\end{equation*}
The total loss is then $\mathcal{J}_{\text{CS}} = \mathcal{J}_{\text{columns}} + \mathcal{J}_{\text{cells}} + \alpha\mathcal{J}_{\text{aggr}}$,
where $\alpha$ is a scaling hyperparameter.

\paragraph{Scalar answer} In this case $y$ is a single scalar $s$ which does not appear in the table (i.e. $C=\emptyset$, e.g., question 2 in Figure \ref{fig:dataset_examples}). This usually corresponds to examples that involve an aggregation over one or more table cells. In this work we handle aggregation operators that correspond to SQL, namely \texttt{COUNT}, \texttt{AVERAGE} and \texttt{SUM}, however our model is not restricted to these. 

For these examples, the table cells that should be selected and the aggregation operator type are not known, as these cannot be directly inferred from the scalar answer $s$. To train the model given this form of supervision one could search offline \cite{dua2019drop,andor2019giving} or online \cite{berant2013freebase, liang2018mapo} for programs (table cells and aggregation) that execute to $s$. In our table parsing setting, the number of spurious programs that execute to the gold scalar answer can grow quickly with the number of table cells (e.g., when $s=5$, each \texttt{COUNT} over any five cells is potentially correct). As with this approach learning can easily fail, we avoid it.

Instead, we make use of a training recipe where no search for correct programs is needed. Our approach results in an end-to-end differentiable training, similar in spirit to \newcite{neelakantan2016neural}. We implement a fully differentiable layer that latently learns the weights for the aggregation prediction layer $p_{\text{a}}(\cdot)$, without explicit supervision for the aggregation type.

Specifically, we recognize that the result of executing each of the supported aggregation operators is a scalar. We then implement a soft differentiable estimation for each operator (Table \ref{tab:operators}), given the token selection probabilities and the table values: $\mbox{compute}(op, p_{\text{s}}, T)$. Given the results for all aggregation operators we then calculate the expected result according to the current model: 
\begin{equation*}\label{eqn:agg}
  s_{\text{pred}} = \sum_{i=1}\hat{p}_{\text{a}}(op_i) \cdot \mbox{compute}(op_i, p_{\text{s}}, T),
\end{equation*}
where $\hat{p}_{\text{a}}(op_i)=\frac{p_{\text{a}}(op_i)}{\sum_{i=1}p_{\text{a}}(op_i)}$ is a probability distribution normalized over aggregation operators excluding \texttt{NONE}. 

\begin{table}[t]
\begin{center}
\resizebox{0.8\columnwidth}{!}{
\begin{tabular}{lc}
\toprule
  $op$ & $\mbox{compute}(op,p_\text{s},T)$ \\
\midrule
 \texttt{COUNT} & $\sum_{c \in T}p^{(c)}_{\text{s}}$  \\
 \texttt{SUM} & $\sum_{c \in T} p^{(c)}_{\text{s}} \cdot T[c]$  \\
 \texttt{AVERAGE} & $\frac{\mbox{compute}(\texttt{SUM},p_\text{s},T)}{\mbox{compute}(\texttt{COUNT},p_\text{s},T)}$  \\
\bottomrule
\end{tabular}}
\end{center}
\caption{Aggregation operators soft implementation. \texttt{AVERAGE} approximation is discussed in Appendix \ref{sec:stochastic_average}. Note that probabilities $p_s^{(c)}$ outside of the column selected by the model are set to 0.}
\label{tab:operators}
\end{table}

We then calculate the scalar answer loss with Huber loss \cite{huber1964robust} given by:
\begin{equation*}
    \mathcal{J}_{\text{scalar}} = 
    \begin{cases} 
      0.5 \cdot a^2 & a \leq \delta \\
      \delta \cdot a - 0.5 \cdot \delta ^2 & \text{otherwise}
   \end{cases}
\end{equation*}
where $a=\mathopen|s_{\text{pred}}-s\mathclose|$, and $\delta$ is a hyperparameter. Like \newcite{neelakantan2016neural}, we find this loss is more stable than the squared loss. In addition, since a scalar answer implies some aggregation operation, we also define an aggregation loss that penalizes the model for assigning probability mass to the \texttt{NONE} class: 
\begin{equation*}
    \mathcal{J}_{\text{aggr}} = -\log(\sum_{i=1}p_{\text{a}}(op_i))
\end{equation*}
The total loss is then $\mathcal{J}_{\text{SA}} = \mathcal{J}_{\text{aggr}} + \beta \mathcal{J}_{\text{scalar}}$,
where $\beta$ is a scaling hyperparameter. As for some examples $\mathcal{J}_{\text{scalar}}$ can be very large, which leads to unstable model updates, we introduce a \textit{cutoff} hyperparameter. Then, for a training example where $\mathcal{J}_{\text{scalar}}>\textit{cutoff}$, we set $\mathcal{J}=0$ to ignore the example entirely, as we noticed this behaviour correlates with outliers. 
In addition, as computation done during training is continuous, while that being done during inference is discrete, we further add a temperature that scales token logits such that $p_{\text{s}}$ would output values closer to binary ones.

\paragraph{Ambiguous answer} A scalar answer $s$ that also appears in the table (thus $C \neq \emptyset$) is ambiguous, as in some cases the question implies aggregation (question 3 in Figure \ref{fig:dataset_examples}), while in other cases a table cell should be predicted (question 4 in Figure \ref{fig:dataset_examples}). Thus, in this case we dynamically let the model choose the supervision (\textit{cell selection} or \textit{scalar answer}) according to its current policy. Concretely, we set the supervision to be of cell selection if $p_{\text{a}} (op_{0}) \ge S$, where $0<S<1$ is a threshold hyperparameter, and the scalar answer supervision otherwise. This follows hard EM \cite{min2019discrete}, as for spurious programs we pick the most probable one according to the current model.

\section{Experiments}
\label{sec:experiments}

\subsection{Datasets}
\label{sec:datasets}

We experiment with the following semantic parsing datasets that reason over single tables (see Table~\ref{tab:dataset_stats}).

\begin{table}[t]
\begin{center}
\scalebox{0.9}{
\begin{tabular}{llll}
\toprule
                          & \textbf{\wikisql{}} & \textbf{\wtq{}} & \textbf{\sqa{}} \\
\midrule
\textbf{Logical Form}     & \cmark     & \xmark & \xmark \\
\textbf{Conversational}   & \xmark     & \xmark & \cmark \\
\textbf{Aggregation}      & \cmark     & \cmark & \xmark \\
\textbf{Examples}         & 80654      & 22033  & 17553  \\
\textbf{Tables}           & 24241      & 2108   & 982    \\
\bottomrule
\end{tabular}
}
\end{center}
\caption{Dataset statistics.}
\label{tab:dataset_stats}
\end{table}

\paragraph{\wtq{} \cite{pasupat2015compositional}} This dataset consists of complex questions on Wikipedia tables. Crowd workers were asked, given a table, to compose a series of complex questions that include comparisons, superlatives, aggregation or arithmetic operation. The questions were then verified by other crowd workers.

\paragraph{\sqa{} \cite{iyyer2017search}} This dataset was constructed by asking crowd workers to decompose a subset of highly compositional questions from \wtq{}, where each resulting decomposed question can be answered by one or more table cells. The final set consists of $6,066$ question sequences ($2.9$ question per sequence on average).

\paragraph{\wikisql{} \cite{zhong2017seq2sql}} This dataset focuses on translating text to SQL. It was constructed by asking crowd workers to paraphrase a template-based question in natural language.
Two other crowd workers were asked to verify the quality of the proposed paraphrases. 

\vspace{3mm}

As our model predicts cell selection or scalar answers, we convert the denotations for each dataset to \la{}question, cell coordinates, scalar answer\ra{} triples. \sqa{} already provides this information (gold cells for each question). For \wikisql{} and \wtq{}, we only use the denotations.
Therefore, we derive cell coordinates by matching the denotations against the table contents. We fill scalar answer information if the denotation contains a single element that can be interpreted as a float, otherwise we set its value to \texttt{NaN}.
We drop examples if there is no scalar answer and the denotation can not be found in the table, or if some denotation matches multiple cells.

\subsection{Experimental Setup}
\label{sec:exp_setup}

We apply the standard BERT tokenizer on questions, table cells and headers, using the same vocabulary of 32k word pieces. Numbers and dates are parsed in a similar way as in the Neural Programmer~\cite{neelakantan2017learning}.

The official evaluation script of \wtq{} and \sqa{} is used to report the denotation accuracy for these datasets. 
For \wikisql{}, we generate the reference answer, aggregation operator and cell coordinates from the reference SQL provided using our own SQL implementation running on the JSON tables.
However, we find that the answer produced by the official \wikisql{} evaluation script is incorrect for approx. $2\%$ of the examples.
Throughout this paper we report accuracies against our reference answers, but we explain the differences and also provide accuracies compared to the official reference answers in Appendix \ref{sec:wikisql_diffs}.

We start pre-training from BERT-Large (see Appendix \ref{sec:wiki_hparams} for hyper-parameters). We find it beneficial to start the pre-training from a pre-trained standard text BERT model (while randomly initializing our additional embeddings), as this enhances convergence on the held-out set.

We run both pre-training and fine-tuning on a setup of 32 Cloud TPU v3 cores with maximum sequence length 512. In this setup pre-training takes around 3 days and fine-tuning around 10 hours for \wikisql{} and \wtq{} and 20 hours for \sqa{} (with the batch sizes from table \ref{tab:hparams}). The resource requirements of our model are essentially the same as BERT-large\footnote{\url{https://github.com/google-research/bert/blob/master/README.md\#out-of-memory-issues}}.

For fine-tuning, we choose hyper-parameters using a black box Bayesian optimizer similar to Google Vizier \cite{vizier} for \wikisql{} and \wtq{}. For \sqa{} we use grid-search. We discuss the details in Appendix \ref{sec:wiki_hparams}.

\subsection{Results}

All results report the denotation accuracy for models trained from weak supervision.
We follow \newcite{niven-kao-2019-probing} and report the median for 5 independent runs, as BERT-based models can degenerate.
We present our results for \wikisql and \wtq in Tables \ref{tab:wikisql_results} and \ref{tab:wtq_results} respectively. 
Table \ref{tab:wikisql_results} shows that \ours, trained in the weakly supervised setting, achieves close to state-of-the-art performance for \wikisql ($83.6$ vs $83.9$ \cite{min2019discrete}).
If given the gold aggregation operators and selected cell as supervision (extracted from the reference SQL), which accounts as full supervision to \ours, the model achieves $86.4$.
Unlike the full SQL queries, this supervision can be annotated by non-experts.

For \wtq the model trained only from the original training data reaches $42.6$ which surpass similar approaches \cite{neelakantan2016neural}. 
When we pre-train the model on \wikisql or \sqa (which is straight-forward in our setup, as we do not rely on a logical formalism), \ours achieves $48.7$ and $48.8$, respectively.

\begin{table}[t]
\begin{center}
\resizebox{0.85\columnwidth}{!}{
\begin{tabular}{lll}
\toprule
  \textbf{Model} & \textbf{Dev} & \textbf{Test} \\
\midrule
  \citet{liang2018mapo}       &  71.8  & 72.4 \\
  \citet{agarwal2019learning}          &  74.9  & 74.8 \\
  \citet{wang2019learning}     &  79.4  & 79.3 \\ 
  \citet{min2019discrete}     &  84.4  & {\bf 83.9} \\
\midrule
\midrule
  \ours                  &   {\bf 85.1} &  83.6 \\
\midrule
  \ours (fully-supervised)     &   88.0 &  86.4 \\
\bottomrule
\end{tabular}}
\end{center}
\caption{\wikisql denotation accuracy\footnotemark.}
\label{tab:wikisql_results}
\end{table}

\begin{table}[t]
\begin{center}
\resizebox{0.85\columnwidth}{!}{
\begin{tabular}{lr}
\toprule
  \textbf{Model} & 
  \textbf{Test} \\
\midrule
 \citet{pasupat2015compositional} &
 37.1 \\
 \citet{neelakantan2017learning} &
 34.2 \\
 \citet{haug2018neural} &
 34.8 \\
 \citet{zhang2017macro} &
 43.7 \\
\citet{liang2018mapo} & 
 43.1  \\
\citet{dasigi2019iterative} &
 43.9 \\
\citet{agarwal2019learning}&
 44.1  \\
\citet{wang2019learning}&
44.5  \\
\midrule
\midrule
  \ours& 42.6 \\
  \ours (pre-trained on \wikisql{}) & 48.7 \\
  \ours (pre-trained on \sqa{}) & 48.8 \\
\bottomrule
\end{tabular}}
\end{center}
\caption{\wtq denotation accuracy.}
\label{tab:wtq_results}
\end{table}

\footnotetext{As explained in Section \ref{sec:exp_setup}, we report \ours numbers comparing against our own reference answers. Appendix \ref{sec:wikisql_diffs} contains numbers WRT the official \wikisql{} eval script.}

For \sqa, Table \ref{tab:sqa_results} shows that \ours leads to substantial improvements on all metrics: Improving all metrics by at least $11$ points, sequence accuracy from $28.1$ to $40.4$ and average question accuracy from $55.1$ to $67.2$.

\begin{table}[t]
\begin{center}
\resizebox{1.0\columnwidth}{!}{
\begin{tabular}{lrrrrr}
\toprule
  \textbf{Model} & \textbf{ALL} & \textbf{SEQ} & \textbf{Q1} & \textbf{Q2} & \textbf{Q3} \\
\midrule
  \citet{pasupat2015compositional}       &  33.2  &  7.7 & 51.4 & 22.2 & 22.3 \\
  \citet{neelakantan2017learning}       &  40.2  & 11.8 & 60.0 & 35.9 & 25.5 \\
  \citet{iyyer2017search}          &  44.7  & 12.8 & 70.4 & 41.1 & 23.6 \\
  \citet{sun2018knowledge}     &  45.6  & 13.2 & 70.3 & 42.6 & 24.8 \\ 
  \citet{muller2019answering}     &  55.1  & 28.1 & 67.2 & 52.7 & 46.8
  \\
\midrule
\midrule
  \ours & \bf 67.2  & \bf 40.4 & \bf 78.2 & \bf 66.0 & \bf 59.7 \\
\bottomrule
\end{tabular}}
\end{center}
\caption{\sqa{} test results. ALL is the average question accuracy, SEQ the sequence accuracy, and QX, the accuracy of the X'th question in a sequence.}
\label{tab:sqa_results}
\end{table}

\begin{table}[t]
\begin{center}
\scalebox{0.7}{
\begin{tabular}{lrrrrrr}
\toprule
  & \multicolumn{2}{c}{\textbf{\sqa} (SEQ)} &  \multicolumn{2}{c}{\textbf{\wikisql{}}} & \multicolumn{2}{c}{\textbf{\wtq}} \\
\midrule
all                   & 39.0 &            & 84.7 &         & 29.0 &       \\
-pos                  & 36.7 & \textcolor{myred}{-2.3}       & 82.9 &  \textcolor{myred}{-1.8} & 25.3  &  \textcolor{myred}{-3.7} \\
-ranks                & 34.4 & \textcolor{myred}{-4.6}       & 84.1 &  \textcolor{myred}{-0.6} & 30.7  &  \textcolor{mygreen}{+1.8} \\
-\{cols,rows\}        & 19.6 & \textcolor{myred}{-19.4}      & 74.1 & \textcolor{myred}{-10.6} & 17.3  & \textcolor{myred}{-11.6} \\
-table pre-training   & 26.5 & \textcolor{myred}{-12.5}      & 80.8 &  \textcolor{myred}{-3.9} & 17.9  & \textcolor{myred}{-11.1} \\   
-aggregation          &    - &                             & 82.6 &  \textcolor{myred}{-2.1} & 23.1  &  \textcolor{myred}{-5.9} \\
\bottomrule
\end{tabular}
}
\end{center}
\caption{Dev accuracy 
with different embeddings removed from the full model: positional (pos), numeric ranks (ranks), column (cols) and row (rows). The model without table pre-training was initialized from the original BERT model pre-trained on text only.
The model without aggregation is only trained with the cell selection loss.}
\label{tab:abl}
\end{table}

\paragraph{Model ablations} 

Table \ref{tab:abl} shows an ablation study on our different embeddings.
To this end we pre-train and fine-tune models with different features.
As pre-training is expensive we limit it to $200,000$ steps.
For all datasets we see that pre-training on tables and column and row embeddings are the most important.
Positional and rank embeddings are also improving the quality but to a lesser extent.

We additionally find that when removing the scalar answer and aggregation losses (i.e., setting $\mathcal{J}_{\text{SA}=0}$) from \ours, accuracy drops for both datasets. For \wtq, we observe a substantial drop in performance from $29.0$ to $23.1$ when removing aggregation. For \wikisql{} performance drops from $84.7$ to $82.6$.
The relatively small decrease for \wikisql{} can be explained by the fact that most examples do not need aggregation to be answered.
In principle, $17\%$ of the examples of the dev set have an aggregation (\texttt{SUM}, \texttt{AVERAGE} or \texttt{COUNT}), however, 
for all types we find that for more than $98\%$ of the examples the aggregation is only applied to one or no cells.
In the case of \texttt{SUM} and \texttt{AVERAGE}, this means that most examples can be answered by selecting one or no cells from the table.
For \texttt{COUNT} the model without aggregation operators achieves $28.2$ accuracy (by selecting $0$ or $1$ from the table) vs. $66.5$ for the model with aggregation.
Note that $0$ and $1$ are often found in a special index column. 
These properties of \wikisql{} make it challenging for the model to decide whether to apply aggregation or not.
For \wtq on the other hand, we observe a substantial drop in performance from $29.0$ to $23.1$ when removing aggregation.

\paragraph{Qualitative Analysis on \wtq}
\label{sec:error_analysis}

We manually analyze $200$ dev set predictions made by \ours on \wtq. For correct predictions via an aggregation, we inspect the selected cells to see if they match the ground truth. We find that $96\%$ of the correct aggregation predictions where also correct in terms of the cells selected. We further find that $14\%$ of the correct aggregation predictions had only one cell, and could potentially be achieved by cell selection, with no aggregation.

We also perform an error analysis and identify the following exclusive salient phenomena:
\begin{enumerate*}[label={(\roman*)},font=\bfseries\em]
\item $12\%$ are ambiguous (\emph{``Name at least two labels that released the group's albums.''}), have wrong labels or missing information ;
\item $10\%$ of the cases require complex temporal comparisons which could also not be parsed with a rich formalism such as SQL (\emph{``what country had the most cities founded in the 1830's?''}) ;
\item in $16\%$ of the cases the gold denotation has a textual value that does not appear in the table, thus it could not be predicted without performing string operations over cell values ;
\item on $10\%$, the table is too big to fit in $512$ tokens ;
\item on $13\%$ of the cases \ours selected no cells, which suggests introducing penalties for this behaviour ;
\item on $2\%$ of the cases, the answer is the difference between scalars, so it is outside of the model capabilities (\emph{``how long did anne churchill/spencer live?''}) ;
\item the other $37\%$ of the cases could not be classified to a particular phenomenon.
\end{enumerate*}

\paragraph{Pre-training Analysis} In order to understand what \ours learns during pre-training we analyze its performance on 10,000 held-out examples. We split the data such that the tables in the held-out data do not occur in the training data.
\begin{table}
\begin{center}
\scalebox{0.9}{
\begin{tabular}{lrrrr}
\toprule
  & \textbf{all} & \textbf{text} & \textbf{header} & \textbf{cell} \\
\midrule
\textbf{all}    & 71.4 & 68.8 & 96.6 & 63.4 \\
\textbf{word}   & 74.1 & 69.7 & 96.9 & 66.6 \\
\textbf{number} & 53.9 & 51.7 & 83.6 & 53.2 \\
\bottomrule
\end{tabular}
}
\end{center}
\caption{Mask LM accuracy on held-out data, when the target word piece is located in the text, table header, cell or anywhere (all) and the target is anything, a word or number.}
\label{tab:pretrain}
\end{table}
Table \ref{tab:pretrain} shows the accuracy of masked word pieces 
of different types and in different locations. We find that average accuracy across position is relatively high (71.4).
Predicting tokens in the header of the table is easiest (96.6), probably because many Wikipedia articles use instances of the same
kind of table. Predicting word pieces in cells is a bit harder (63.4) than predicting pieces in the text (68.8). 
The biggest differences can be observed when comparing predicting words (74.1) and numbers (53.9).
This is expected since numbers are very specific and often hard to generalize. The soft-accuracy metric and example (Appendix \ref{sec:number_pretrain_example}) demonstrate, however, that the model is relatively good at predicting numbers that are at least close to the target.

\paragraph{Limitations}

\ours handles single tables as context, which are able to fit in memory. Thus, our model would fail to capture very large tables, or databases that contain multiple tables. In this case, the table(s) could be compressed or filtered, such that only relevant content would be encoded, which we leave for future work.

In addition, although \ours can parse compositional structures (e.g., question 2 in Figure \ref{fig:dataset_examples}), its expressivity is limited to a form of an aggregation over a subset of table cells. Thus, structures with multiple aggregations such as \textit{``number of actors with an average rating higher than 4''} could not be handled correctly. Despite this limitation, \ours succeeds in parsing three different datasets, and we did not encounter this kind of errors in Section \ref{sec:error_analysis}. This suggests that the majority of examples in semantic parsing datasets are limited in their compositionality.

\section{Related Work}

Semantic parsing models are mostly trained to produce gold logical forms using an encoder-decoder approach \cite{jia2016recombination,dong2016logical}.
To reduce the burden in collecting full logical forms, models are typically trained from weak supervision in the form of denotations. These are used to guide the search for correct logical forms \cite{clarke10world,liang11dcs}. 

Other works suggested end-to-end differentiable models that train from weak supervision, but do not explicitly generate logical forms. \newcite{neelakantan2016neural} proposed a complex model that sequentially predicts symbolic operations over table segments that are all explicitly predefined by the authors, while \newcite{yin-etal-2016-neural} proposed a similar model where the operations themselves are learned during training. \newcite{muller2019answering} proposed a model that selects table cells, where the table and question are represented as a Graph Neural Network, however their model can not predict aggregations over table cells. 
\newcite{Cho2018AdversarialTA} proposed a supervised model that predicts the relevant rows, column and aggregation operation sequentially.
In our work, we propose a model that follow this line of work, with a simpler architecture than past models (as the model is a single encoder that performs computation for many operations implicitly) and more coverage (as we support aggregation operators over selected cells).

Finally, pre-training methods have been designed with different training objectives, including language modeling \cite{dai2015semi,peters2018elmo,radford2018improving} and masked language modeling \cite{devlin2018BERT, lample2019cross}. These methods dramatically boost the performance of natural language understanding models \cite[\emph{inter alia}]{peters2018elmo}. Recently, several works extended BERT for visual question answering, by pre-training over text-image pairs while masking different regions in the image \cite{tan2019lxmert,lu2019vilbert}. As for tables, \newcite{TabFact} experimented with rendering a table into natural language so that it can be handled with a pre-trained BERT model. In our work we extend masked language modeling for table representations, by masking table cells or text segments.

\section{Conclusion}

In this paper we presented \ours, a model for question answering over tables that avoids generating logical forms. We showed that \ours effectively pre-trains over large scale data of text-table pairs and successfully restores masked words and table cells. We additionally showed that the model can fine-tune on semantic parsing datasets, only using weak supervision, with an end-to-end differentiable recipe. Results show that \ours achieves better or competitive results in comparison to state-of-the-art semantic parsers.

In future work we aim to extend the model to represent a database with multiple tables as context, and to effectively handle large tables.

\section{Acknowledgments}

We would like to thank Yasemin Altun, Srini Narayanan, Slav Petrov, William Cohen, Massimo Nicosia, Syrine Krichene, Jordan Boyd-Graber and the anonymous reviewers for their constructive feedback, useful comments and suggestions. This work was completed in partial fulfillment for the PhD degree of the first author, which was also supported by a Google PhD fellowship.

\bibliography{references}

\begin{thebibliography}{47}
\expandafter\ifx\csname natexlab\endcsname\relax\def\natexlab#1{#1}\fi

\bibitem[{Agarwal et~al.(2019)Agarwal, Liang, Schuurmans, and
  Norouzi}]{agarwal2019learning}
R.~Agarwal, C.~Liang, D.~Schuurmans, and M.~Norouzi. 2019.
\newblock Learning to generalize from sparse and underspecified rewards.
\newblock \emph{arXiv preprint arXiv:1902.07198}.

\bibitem[{Andor et~al.(2016)Andor, Alberti, Weiss, Severyn, Presta, Ganchev,
  Petrov, and Collins}]{andor2016globally}
D.~Andor, C.~Alberti, D.~Weiss, A.~Severyn, A.~Presta, K.~Ganchev, S.~Petrov,
  and M.~Collins. 2016.
\newblock Globally normalized transition-based neural networks.
\newblock \emph{arXiv preprint arXiv:1603.06042}.

\bibitem[{Andor et~al.(2019)Andor, He, Lee, and Pitler}]{andor2019giving}
Daniel Andor, Luheng He, Kenton Lee, and Emily Pitler. 2019.
\newblock Giving bert a calculator: Finding operations and arguments with
  reading comprehension.
\newblock \emph{arXiv preprint arXiv:1909.00109}.

\bibitem[{Artzi and Zettlemoyer(2013)}]{artzi2013weakly}
Y.~Artzi and L.~Zettlemoyer. 2013.
\newblock Weakly supervised learning of semantic parsers for mapping
  instructions to actions.
\newblock \emph{Transactions of the Association for Computational Linguistics
  (TACL)}, 1:49--62.

\bibitem[{Berant et~al.(2013)Berant, Chou, Frostig, and
  Liang}]{berant2013freebase}
J.~Berant, A.~Chou, R.~Frostig, and P.~Liang. 2013.
\newblock Semantic parsing on {F}reebase from question-answer pairs.
\newblock In \emph{Empirical Methods in Natural Language Processing (EMNLP)}.

\bibitem[{Chen et~al.(2019)Chen, Wang, Chen, Zhang, Wang, Li, Zhou, and
  Wang}]{TabFact}
Wenhu Chen, Hongmin Wang, Jianshu Chen, Yunkai Zhang, Hong Wang, Shiyang Li,
  Xiyou Zhou, and William~Yang Wang. 2019.
\newblock Tabfact: A large-scale dataset for table-based fact verification.
\newblock \emph{ArXiv}, abs/1909.02164.

\bibitem[{Cho et~al.(2018)Cho, Amplayo, won Hwang, and
  Park}]{Cho2018AdversarialTA}
Minseok Cho, Reinald~Kim Amplayo, Seung won Hwang, and Jonghyuck Park. 2018.
\newblock Adversarial tableqa: Attention supervision for question answering on
  tables.
\newblock In \emph{ACML}.

\bibitem[{Clarke et~al.(2010)Clarke, Goldwasser, Chang, and
  Roth}]{clarke10world}
J.~Clarke, D.~Goldwasser, M.~Chang, and D.~Roth. 2010.
\newblock Driving semantic parsing from the world's response.
\newblock In \emph{Computational Natural Language Learning (CoNLL)}, pages
  18--27.

\bibitem[{Dai and Le(2015)}]{dai2015semi}
A.~M. Dai and Q.~V. Le. 2015.
\newblock Semi-supervised sequence learning.
\newblock In \emph{Advances in Neural Information Processing Systems
  (NeurIPS)}.

\bibitem[{Dasigi et~al.(2019)Dasigi, Gardner, Murty, Zettlemoyer, and
  Hovy}]{dasigi2019iterative}
Pradeep Dasigi, Matt Gardner, Shikhar Murty, Luke Zettlemoyer, and Eduard Hovy.
  2019.
\newblock Iterative search for weakly supervised semantic parsing.
\newblock In \emph{Proceedings of the 2019 Conference of the North American
  Chapter of the Association for Computational Linguistics: Human Language
  Technologies, Volume 1 (Long and Short Papers)}, pages 2669--2680.

\bibitem[{Devlin et~al.(2019)Devlin, Chang, Lee, and
  Toutanova}]{devlin2018BERT}
Jacob Devlin, Ming-Wei Chang, Kenton Lee, and Kristina Toutanova. 2019.
\newblock \href {https://doi.org/10.18653/v1/N19-1423} {{BERT}: Pre-training of
  deep bidirectional transformers for language understanding}.
\newblock In \emph{Proceedings of the 2019 Conference of the North {A}merican
  Chapter of the Association for Computational Linguistics: Human Language
  Technologies, Volume 1 (Long and Short Papers)}, pages 4171--4186,
  Minneapolis, Minnesota. Association for Computational Linguistics.

\bibitem[{Dong and Lapata(2016)}]{dong2016logical}
L.~Dong and M.~Lapata. 2016.
\newblock Language to logical form with neural attention.
\newblock In \emph{Association for Computational Linguistics (ACL)}.

\bibitem[{Dua et~al.(2019)Dua, Wang, Dasigi, Stanovsky, Singh, and
  Gardner}]{dua2019drop}
D.~Dua, Y.~Wang, P.~Dasigi, G.~Stanovsky, S.~Singh, and M.~Gardner. 2019.
\newblock {DROP}: A reading comprehension benchmark requiring discrete
  reasoning over paragraphs.
\newblock In \emph{North American Association for Computational Linguistics
  (NAACL)}.

\bibitem[{Golovin et~al.(2017)Golovin, Solnik, Moitra, Kochanski, Karro, and
  Sculley}]{vizier}
Daniel Golovin, Benjamin Solnik, Subhodeep Moitra, Greg Kochanski, John~Elliot
  Karro, and D.~Sculley, editors. 2017.
\newblock \href
  {http://www.kdd.org/kdd2017/papers/view/google-vizier-a-service-for-black-box-optimization}
  {\emph{Google Vizier: A Service for Black-Box Optimization}}.

\bibitem[{Guu et~al.(2017)Guu, Pasupat, Liu, and Liang}]{guu2017bridging}
K.~Guu, P.~Pasupat, E.~Z. Liu, and P.~Liang. 2017.
\newblock From language to programs: Bridging reinforcement learning and
  maximum marginal likelihood.
\newblock In \emph{Association for Computational Linguistics (ACL)}.

\bibitem[{Haug et~al.(2018)Haug, Ganea, and Grnarova}]{haug2018neural}
T.~Haug, O.~Ganea, and P.~Grnarova. 2018.
\newblock Neural multi-step reasoning for question answering on semi-structured
  tables.
\newblock In \emph{European Conference on Information Retrieval}.

\bibitem[{Herzig and Berant(2017)}]{herzig2017multi}
J.~Herzig and J.~Berant. 2017.
\newblock Neural semantic parsing over multiple knowledge-bases.
\newblock In \emph{Association for Computational Linguistics (ACL)}.

\bibitem[{Huber(1964)}]{huber1964robust}
P.~J. Huber. 1964.
\newblock Robust estimation of a location parameter.
\newblock \emph{The Annals of Mathematical Statistics}, 35(1):73--101.

\bibitem[{Hwang et~al.(2019)Hwang, Yim, Park, and Seo}]{hwang2019comprehensive}
Wonseok Hwang, Jinyeung Yim, Seunghyun Park, and Minjoon Seo. 2019.
\newblock A comprehensive exploration on wikisql with table-aware word
  contextualization.
\newblock \emph{arXiv preprint arXiv:1902.01069}.

\bibitem[{Iyer et~al.(2017)Iyer, Konstas, Cheung, Krishnamurthy, and
  Zettlemoyer}]{iyer2017neural}
S.~Iyer, I.~Konstas, A.~Cheung, J.~Krishnamurthy, and L.~Zettlemoyer. 2017.
\newblock Learning a neural semantic parser from user feedback.
\newblock In \emph{Association for Computational Linguistics (ACL)}.

\bibitem[{Iyyer et~al.(2017)Iyyer, Yih, and Chang}]{iyyer2017search}
M.~Iyyer, W.~Yih, and M.~Chang. 2017.
\newblock Search-based neural structured learning for sequential question
  answering.
\newblock In \emph{Association for Computational Linguistics (ACL)}.

\bibitem[{Jia and Liang(2016)}]{jia2016recombination}
R.~Jia and P.~Liang. 2016.
\newblock Data recombination for neural semantic parsing.
\newblock In \emph{Association for Computational Linguistics (ACL)}.

\bibitem[{Lafferty et~al.(2001)Lafferty, McCallum, and Pereira}]{lafferty01crf}
J.~Lafferty, A.~McCallum, and F.~Pereira. 2001.
\newblock Conditional random fields: Probabilistic models for segmenting and
  labeling data.
\newblock In \emph{International Conference on Machine Learning (ICML)}, pages
  282--289.

\bibitem[{Lample and Conneau(2019)}]{lample2019cross}
Guillaume Lample and Alexis Conneau. 2019.
\newblock Cross-lingual language model pretraining.
\newblock \emph{arXiv preprint arXiv:1901.07291}.

\bibitem[{Lawrence and Riezler(2018)}]{lawrence-riezler-2018-improving}
Carolin Lawrence and Stefan Riezler. 2018.
\newblock \href {https://www.aclweb.org/anthology/P18-1169} {Improving a neural
  semantic parser by counterfactual learning from human bandit feedback}.
\newblock In \emph{Proceedings of the 56th Annual Meeting of the Association
  for Computational Linguistics (Volume 1: Long Papers)}, pages 1820--1830,
  Melbourne, Australia. Association for Computational Linguistics.

\bibitem[{Liang et~al.(2018)Liang, Norouzi, Berant, Le, and
  Lao}]{liang2018mapo}
C.~Liang, M.~Norouzi, J.~Berant, Q.~Le, and N.~Lao. 2018.
\newblock Memory augmented policy optimization for program synthesis with
  generalization.
\newblock In \emph{Advances in Neural Information Processing Systems
  (NeurIPS)}.

\bibitem[{Liang et~al.(2011)Liang, Jordan, and Klein}]{liang11dcs}
P.~Liang, M.~I. Jordan, and D.~Klein. 2011.
\newblock Learning dependency-based compositional semantics.
\newblock In \emph{Association for Computational Linguistics (ACL)}, pages
  590--599.

\bibitem[{Liu et~al.(2019)Liu, Ott, Goyal, Du, Joshi, Chen, Levy, Lewis,
  Zettlemoyer, and Stoyanov}]{roberta2019}
Yinhan Liu, Myle Ott, Naman Goyal, Jingfei Du, Mandar Joshi, Danqi Chen, Omer
  Levy, Mike Lewis, Luke Zettlemoyer, and Veselin Stoyanov. 2019.
\newblock \href {http://arxiv.org/abs/1907.11692} {Roberta: {A} robustly
  optimized {BERT} pretraining approach}.
\newblock \emph{CoRR}, abs/1907.11692.

\bibitem[{Lu et~al.(2019)Lu, Batra, Parikh, and Lee}]{lu2019vilbert}
Jiasen Lu, Dhruv Batra, Devi Parikh, and Stefan Lee. 2019.
\newblock Vilbert: Pretraining task-agnostic visiolinguistic representations
  for vision-and-language tasks.
\newblock \emph{arXiv preprint arXiv:1908.02265}.

\bibitem[{Min et~al.(2019)Min, Chen, Hajishirzi, and
  Zettlemoyer}]{min2019discrete}
Sewon Min, Danqi Chen, Hannaneh Hajishirzi, and Luke Zettlemoyer. 2019.
\newblock \href {https://doi.org/10.18653/v1/D19-1284} {A discrete hard {EM}
  approach for weakly supervised question answering}.
\newblock In \emph{Proceedings of the 2019 Conference on Empirical Methods in
  Natural Language Processing and the 9th International Joint Conference on
  Natural Language Processing (EMNLP-IJCNLP)}, pages 2851--2864, Hong Kong,
  China. Association for Computational Linguistics.

\bibitem[{Muhlgay et~al.(2019)Muhlgay, Herzig, and Berant}]{muhlgay2019value}
D.~Muhlgay, J.~Herzig, and J.~Berant. 2019.
\newblock Value-based search in execution space for mapping instructions to
  programs.
\newblock In \emph{North American Association for Computational Linguistics
  (NAACL)}.

\bibitem[{M{\"u}ller et~al.(2019)M{\"u}ller, Piccinno, Nicosia, Shaw, and
  Altun}]{muller2019answering}
Thomas M{\"u}ller, Francesco Piccinno, Massimo Nicosia, Peter Shaw, and Yasemin
  Altun. 2019.
\newblock Answering conversational questions on structured data without logical
  forms.
\newblock \emph{arXiv preprint arXiv:1908.11787}.

\bibitem[{Neelakantan et~al.(2017)Neelakantan, Le, Abadi, McCallum, and
  Amodei}]{neelakantan2017learning}
A.~Neelakantan, Q.~V. Le, M.~Abadi, A.~McCallum, and D.~Amodei. 2017.
\newblock Learning a natural language interface with neural programmer.
\newblock In \emph{International Conference on Learning Representations
  (ICLR)}.

\bibitem[{Neelakantan et~al.(2015)Neelakantan, Le, and
  Sutskever}]{neelakantan2016neural}
Arvind Neelakantan, Quoc~V. Le, and Ilya Sutskever. 2015.
\newblock Neural programmer: Inducing latent programs with gradient descent.
\newblock \emph{CoRR}, abs/1511.04834.

\bibitem[{Niven and Kao(2019)}]{niven-kao-2019-probing}
Timothy Niven and Hung-Yu Kao. 2019.
\newblock \href {https://doi.org/10.18653/v1/P19-1459} {Probing neural network
  comprehension of natural language arguments}.
\newblock In \emph{Proceedings of the 57th Annual Meeting of the Association
  for Computational Linguistics}, pages 4658--4664, Florence, Italy.
  Association for Computational Linguistics.

\bibitem[{Pasupat and Liang(2015)}]{pasupat2015compositional}
Panupong Pasupat and Percy Liang. 2015.
\newblock \href {https://doi.org/10.3115/v1/P15-1142} {Compositional semantic
  parsing on semi-structured tables}.
\newblock In \emph{Proceedings of the 53rd Annual Meeting of the Association
  for Computational Linguistics and the 7th International Joint Conference on
  Natural Language Processing (Volume 1: Long Papers)}, pages 1470--1480,
  Beijing, China. Association for Computational Linguistics.

\bibitem[{Peters et~al.(2018)Peters, Neumann, Iyyer, Gardner, Clark, Lee, and
  Zettlemoyer}]{peters2018elmo}
M.~E. Peters, M.~Neumann, M.~Iyyer, M.~Gardner, C.~Clark, K.~Lee, and
  L.~Zettlemoyer. 2018.
\newblock Deep contextualized word representations.
\newblock In \emph{North American Association for Computational Linguistics
  (NAACL)}.

\bibitem[{Radford et~al.(2018)Radford, Narasimhan, Salimans, and
  Sutskever}]{radford2018improving}
A.~Radford, K.~Narasimhan, T.~Salimans, and I.~Sutskever. 2018.
\newblock Improving language understanding by generative pre-training.
\newblock Technical report, OpenAI.

\bibitem[{Su and Yan(2017)}]{su2017cross}
Y.~Su and X.~Yan. 2017.
\newblock Cross-domain semantic parsing via paraphrasing.
\newblock In \emph{Empirical Methods in Natural Language Processing (EMNLP)}.

\bibitem[{Sun et~al.(2018)Sun, Tang, Duan, Xu, Feng, and
  Qin}]{sun2018knowledge}
Yibo Sun, Duyu Tang, Nan Duan, Jingjing Xu, Xiaocheng Feng, and Bing Qin. 2018.
\newblock Knowledge-aware conversational semantic parsing over web tables.
\newblock \emph{arXiv preprint arXiv:1809.04271}.

\bibitem[{Tan and Bansal(2019)}]{tan2019lxmert}
Hao Tan and Mohit Bansal. 2019.
\newblock Lxmert: Learning cross-modality encoder representations from
  transformers.
\newblock In \emph{Proceedings of the 2019 Conference on Empirical Methods in
  Natural Language Processing}.

\bibitem[{Wallace et~al.(2019)Wallace, Wang, Li, Singh, and
  Gardner}]{wallace2019nlp}
Eric Wallace, Yizhong Wang, Sujian Li, Sameer Singh, and Matt Gardner. 2019.
\newblock Do nlp models know numbers? probing numeracy in embeddings.
\newblock \emph{arXiv preprint arXiv:1909.07940}.

\bibitem[{Wang et~al.(2019)Wang, Titov, and Lapata}]{wang2019learning}
Bailin Wang, Ivan Titov, and Mirella Lapata. 2019.
\newblock \href {https://doi.org/10.18653/v1/D19-1391} {Learning semantic
  parsers from denotations with latent structured alignments and abstract
  programs}.
\newblock In \emph{Proceedings of the 2019 Conference on Empirical Methods in
  Natural Language Processing and the 9th International Joint Conference on
  Natural Language Processing (EMNLP-IJCNLP)}, pages 3772--3783, Hong Kong,
  China. Association for Computational Linguistics.

\bibitem[{Wang et~al.(2015)Wang, Berant, and Liang}]{wang2015overnight}
Y.~Wang, J.~Berant, and P.~Liang. 2015.
\newblock Building a semantic parser overnight.
\newblock In \emph{Association for Computational Linguistics (ACL)}.

\bibitem[{Yin et~al.(2016)Yin, Lu, Li, and Ben}]{yin-etal-2016-neural}
Pengcheng Yin, Zhengdong Lu, Hang Li, and Kao Ben. 2016.
\newblock \href {https://doi.org/10.18653/v1/W16-0105} {Neural enquirer:
  Learning to query tables in natural language}.
\newblock In \emph{Proceedings of the Workshop on Human-Computer Question
  Answering}, pages 29--35, San Diego, California. Association for
  Computational Linguistics.

\bibitem[{Zhang et~al.(2017)Zhang, Pasupat, and Liang}]{zhang2017macro}
Y.~Zhang, P.~Pasupat, and P.~Liang. 2017.
\newblock Macro grammars and holistic triggering for efficient semantic
  parsing.
\newblock In \emph{Empirical Methods in Natural Language Processing (EMNLP)}.

\bibitem[{Zhong et~al.(2017)Zhong, Xiong, and Socher}]{zhong2017seq2sql}
Victor Zhong, Caiming Xiong, and Richard Socher. 2017.
\newblock \href {http://arxiv.org/abs/1709.00103} {Seq2sql: Generating
  structured queries from natural language using reinforcement learning}.
\newblock \emph{CoRR}, abs/1709.00103.

\end{thebibliography}
\bibliographystyle{acl_natbib}

\newpage
\newpage

\newpage

\appendix

\section{\wikisql Execution Errors}
\label{sec:wikisql_diffs}

\begin{table*}
\begin{center}
\scalebox{0.85}{
\begin{tabular}{llllll}
\toprule
\texttt{col0} & \texttt{col1} & \texttt{col2} & \texttt{col3} & \texttt{col4} & \texttt{col5} \\
\textbf{Home team} & \textbf{Home team score} & \textbf{Away team} & \textbf{Away team score} & \textbf{Venue} & \textbf{Crowd} \\
\midrule
geelong&18.17 (125)&hawthorn&6.7 (43)&corio oval&9,000\\
footscray&8.18 (66)&south melbourne&11.18 (84)&western oval&12,500\\
fitzroy&11.5 (71)&richmond&8.12 (60)&brunswick street oval&14,000\\
north melbourne&6.12 (48)&essendon&14.11 (95)&arden street oval&8,000\\
st kilda&14.7 (91)&collingwood&17.13 (115)&junction oval&16,000\\
melbourne&12.11 (83)&carlton&11.11 (77)&mcg&31,481\\
\bottomrule
\end{tabular}
}
\scalebox{0.7}{
\begin{tabular}{ll}
\\
\toprule
Question & What was the away team's score when the crowd at Arden Street Oval was larger than 31,481? \\
SQL Query&

\texttt{SELECT col3 AS result FROM table\_2\_10767641\_15}
\\
&
\texttt{WHERE col5 > 31481.0 AND col4 = "arden street oval"}
\\
\wikisql answer & \texttt{["14.11 (95)"]} \\
Our answer & \texttt{[]} \\
\\
\midrule
Question & What was the sum of the crowds at Western Oval? \\
SQL Query& 
\texttt{SELECT SUM(col5) AS result FROM table\_2\_10767641\_15}
\\
& 
\texttt{WHERE col4 = "western oval"}
\\
\wikisql answer & \texttt{[12.0]} \\
Our answer & \texttt{[12500.0]} \\
\\
\bottomrule
\end{tabular}
}
\end{center}
\caption{Table ``2-10767641-15'' from \wikisql. ``\texttt{col6}'' was removed. The ``{Crowd}'' column is of type ``\texttt{REAL}'' but the cell values are actually stored as ``\texttt{TEXT}''.
Below we have two questions from the training set with the answer that is produced by the \wikisql evaluation script and the answer we derive.}
\label{tabl:wikisql_diff}
\end{table*}

In some tables, \wikisql contains ``\texttt{REAL}'' numbers stored in ``\texttt{TEXT}'' format. This leads to incorrect results for some of the comparison and aggregation examples. 
These errors in the \wikisql{} execution accuracy penalize systems that do their own execution (rather then producing an SQL query).
Table \ref{tabl:wikisql_diff} shows two examples where our result derivation and the one used by \wikisql differ because the numbers in the ``Crowd'' (\texttt{col5}) column are not represented as numbers in the respective SQL table.
Table \ref{tab:wiki_off_dev} and \ref{tab:wiki_off_test} contain accuracies compared against the official and our answers.

\begin{table}[H]
\begin{center}
\resizebox{0.85\columnwidth}{!}{
\begin{tabular}{lrrrrr}
\toprule
  \textbf{Model} & \textbf{\wikisql} & \textbf{\ours} \\
  \midrule
  \ours (no answer loss) &   81.2 & 82.5 \\
  \ours                  &   83.9 & 85.1 \\
  \ours (supervised)     &   86.6 & 88.0 \\
\bottomrule
\end{tabular}}
\end{center}
\caption{\wikisql development denotation accuracy.}
\label{tab:wiki_off_dev}
\end{table}

\begin{table}[H]
\begin{center}
\resizebox{0.85\columnwidth}{!}{
\begin{tabular}{lrrrrr}
\toprule
  \textbf{Model} & \textbf{\wikisql} & \textbf{\ours} \\
  \midrule
  \ours (no answer loss) &   80.1 & 81.2 \\
  \ours                  &   82.4 & 83.6 \\
  \ours (supervised)     &   85.2 & 86.4 \\
\bottomrule
\end{tabular}}
\end{center}
\caption{\wikisql test denotation accuracy.}
\label{tab:wiki_off_test}
\end{table}

\section{Hyperparameters}
\label{sec:wiki_hparams}

\begin{table}[H]
\begin{center}
\resizebox{0.85\columnwidth}{!}{
    \begin{tabular}{lll}
    \toprule
    \textbf{Parameter} & \textbf{Values} & \textbf{Scale} \\
    \midrule
    Learning rate & (1e-5, 3e-3) & Log \\
    Warmup ratio & (0.0, 0.2) & Linear \\
    Temperature & (0.1, 1) & Linear \\
    Answer loss cutoff & (0.1, 10,000) & Log \\
    Huber loss delta & (0.1, 10,000) & Log \\
    Cell selection preference & (0, 1) & Linear \\
    Reset cell selection weights & [0, 1] & Discrete \\
    \bottomrule
    \end{tabular}
}
\end{center}
\caption{Hyper-parameters for \wikisql and \wtq. Values are constrained to either a range $(a, b)$ or a list $[a, b, c, \ldots]$.}
\end{table}

\begin{table}[H]
\begin{center}
\resizebox{0.85\columnwidth}{!}{
    \begin{tabular}{lrrrr}
    \toprule
    \textbf{Parameter}                   & \textbf{\textsc{PRETRAIN}} &         \textbf{\sqa} & \textbf{\wikisql} & \textbf{\wtq} \\
    \midrule
    Training Steps              & 1,000,000 &        200,000 &         50,000 & 50,000 \\
    Learning rate               &      5e-5 & 1.25e-5 &     6.17164e-5 &    1.93581e-5 \\
    Warmup ratio                &      0.01 &     0.2 &       0.142400 &      0.128960 \\
    Temperature                 &           &     1.0 &       0.107515 &     0.0352513 \\
    Answer loss cutoff          &           &         &       0.185567 &      0.664694 \\
    Huber loss delta            &           &         &        1265.74 &      0.121194 \\
    Cell selection preference   &           &         &       0.611754 &      0.207951 \\
    Batch size                  &      512  &     128 &            512 &           512 \\
    Gradient clipping           &           &         &             10 &            10 \\
    Select one column           &           &       1 &              0 &             1 \\
    Reset cell selection weights&           &       0 &              0 &             1 \\
    \bottomrule
    \end{tabular}
}
\end{center}
\caption{Optimal hyper-parameters found for pretraining (\textsc{PRETRAIN}), \sqa, \wikisql and \wtq.}
\label{tab:hparams}
\end{table}

\section{Pre-training Example}
\label{sec:number_pretrain_example}

In order to better understand how well the model predicts numbers,
we relax our accuracy measure to a soft form of accuracy: 

\begin{small}
\begin{equation*}
acc(x, y) = \begin{cases} 1 & \mbox{if } x = y \\
 0 & \mbox{if $x$ or $y$ is not a number}\\
 1.0 - \frac{|x - y|}{\max(x,y)} & \mbox{else}
\end{cases}
\end{equation*}
\end{small}

With this soft metric we get an overall accuracy of 74.5 (instead of 71.4) and an accuracy of 80.5 (instead of 53.9) for numbers. Showing that the model is pretty good at guessing numbers that are at least close to the target. The following example demonstrates this:

\begin{table}[H]
\begin{center}
\scalebox{0.7}{
\begin{tabular}{lcccccccc}
 \textbf{Team}         & \textbf{Pld} & \textbf{W} & \textbf{D} & \textbf{L} & \textbf{PF} & \textbf{PA} & \textbf{PD} & \textbf{Pts} \\
\midrule
South Korea   & 2   & 1 & 1 & 0 & 33 & 22 & 11  & 5   \\
     Spain    & 2   & 1 & \la\textbf{\textcolor{dgreen}{1}}\ra & \la\textbf{\textcolor{dgreen}{0}}\ra & 31 & 24 & 7   & 5   \\
     Zimbabwe & 2   & 0 & 0 & 2 & 22 & \la\textbf{\textcolor{red}{43}},\textbf{\textcolor{dgreen}{40}}\ra & - \la\textbf{\textcolor{red}{19}},\textbf{\textcolor{dgreen}{18}}\ra & 2   \\
\end{tabular}}
\end{center}
\caption{Table example from the Wikipedia page describing the 1997 Rugby World Cup Sevens. \la{}x\ra{} marks
a correct prediction and \la{}x,y\ra{} an incorrect prediction.}

\end{table}

In the example, the model correctly restores the Draw (D) and Loss (L) numbers for Spain. It fails to restore the Points For (PF) and Points Against (PA) for Zimbabwe, but gives close estimates. Note that the model also does not produce completely consistent results for each row we should have $\operatorname{PA} + \operatorname{PD} = \operatorname{PF}$ and the column sums of PF and PA should equal.

\section{The average of stochastic sets}
\label{sec:stochastic_average}

Our approach to estimate aggregates of cells in the table operates directly on latent conditionally independent Bernoulli variables $G_c \sim \text{Bern}(p_c)$ that indicate whether each cell is included in the aggregation and a latent categorical variable that indicates the chosen aggregation operation \emph{op}: \texttt{AVERAGE}, \texttt{SUM} or \texttt{COUNT}. Given $G_c$ and the table values $T$ we can define a random subset $S\subseteq T$ where $p_c = P(c\in S)$ for each cell $c\in T$.

The expected value of $\texttt{COUNT}(S) = \sum_c G_c$ can be computed as $\sum_c p_c$ and $\texttt{SUM}(S) = \sum_c G_c T_c$ as $\sum_c p_c T_c$ as described in Table \ref{tab:operators}. For the average however, this is not straight-forward. We will see in what follows that the quotient of the expected sum and the count, which equals the weighed average of $T$ by $p_c$ in general is not the true expected value, which can be written as:

\begin{equation*}
    \mathbb{E}\left[\frac{\sum G_c T_c}{\sum G_c}\right]
\end{equation*}

This quantity differs from the weighted average, a key difference being that the weighted average is not sensitive to constants scaling all the output probabilities, which could in theory find optima where all the $p_c$ are below $0.5$ for example. By the linearity of the expectation we can write:

\begin{equation*}
    \sum_c T_c \mathbb{E}\left[\frac{G_c}{\sum_j G_j}\right] = \sum_c T_c p_c \mathbb{E}\left[\frac{1}{1 + \sum_{j \neq c} G_j}\right]
\end{equation*}

So it comes down to computing that quantity $Q_c = \mathbb{E}\left[\frac{1}{X_c}\right] = \mathbb{E}\left[\frac{1}{1 + \sum_{j \neq c} G_j}\right]$. 
The key observation is that this is the expectation of a reciprocal of a \emph{Poisson Binomial Distribution} \footnote{\href{https://en.wikipedia.org/wiki/Poisson_binomial_distribution}{wikipedia.org/Poisson\_binomial\_distribution}} (a sum of Bernoulli variables) in the special case where one of the probabilities is $1$.

By using the \emph{Jensen inequality} we get a lower bound on
$Q_c$ as $\frac{1}{\mathbb{E}\left[X_c\right]} = \frac{1}{1+\sum_{j\neq c}p_j}$. 
Note that if instead we used $\frac{1}{\sum_{j} p_j}$ then we recover the weighted average, which is strictly bigger than the lower bound and in general not an upper or lower bound. 
We can get better approximations by computing the \emph{Taylor expansion using the
moments}\footnote{\href{https://en.wikipedia.org/wiki/Taylor_expansions_for_the_moments_of_functions_of_random_variables}{wikipedia.org/Taylor\_expansions\_for\_the\_moments}} of $X_c$ of order $k$:
\begin{align*}
Q_c = \mathbb{E}\left[\frac{1}{X_c}\right] \simeq & \frac{1}{\mathbb{E}\left[X_c\right]}
+ \frac{\text{var}\left[X_c\right]}{\mathbb{E}\left[X_c\right]^3}
+ \cdots +\\
& (-1)^k \frac{\mathbb{E}\left[\left(X_c - \mathbb{E}\left[X_c\right]\right)^k\right]}{\mathbb{E}\left[X_c\right]^{k+1}}
\end{align*}

where $\text{var}\left[X_c\right] = \sum_{j\neq c} p_j (1-p_j)$.

The full form for the zero and second order Taylor approximations are:

\begin{align*}
\texttt{AVERAGE}_0(T, p) &= \sum_c T_c \frac{p_c}{1 + \sum_{j\neq c}p_j}  \\
\texttt{AVERAGE}_2(T, p) &= \sum_c T_c \frac{p_c (1 + \epsilon_c)}{1 + \sum_{j\neq c}p_j} \\
\text{with}\: \epsilon_c &= \frac{\sum_{j\neq c} p_j(1-p_j)}{(1 + \sum_{j\neq c}p_j)^2}
\end{align*}

The approximations are then easy to write in any tensor computation language and will be differentiable.
In this work we experimented with the zero and second order approximations and found small improvements over the weighted average baseline. It's worth noting that in the dataset the proportion of average examples is very low. We expect this method to be more relevant in the more general setting.

\end{document}